\documentclass[%
superscriptaddress,
longbibliography,
twocolumn,
nofootinbib,
amsmath,amssymb,
prb,
]{revtex4-2}

\usepackage{graphicx}
\usepackage{dcolumn}
\usepackage{bm}
\usepackage{lettrine}

\usepackage[utf8]{inputenc}
\usepackage[T1]{fontenc}
\usepackage{mathptmx}
\usepackage{etoolbox}

\usepackage{xcolor}

\begin{document}


\title{
Observation of ultra-high-$Q$ resonators in the ultrasound via bound states in the continuum}
\author{Mohamed Farhat}
\affiliation{Computer, Electrical, and Mathematical Sciences and Engineering Division, King Abdullah University of Science and Technology (KAUST), Thuwal 23955-6900, Saudi Arabia}

\author{Younes Achaoui}
\affiliation{Institut FEMTO-ST, CNRS UMR 6174, University Bourgogne Franche-Comté, 15B Avenue des Montboucons, 25000 Besançon Cedex, France}
\affiliation{Faculté des sciences, Université Moulay Ismail, Meknes, Morocco}

\author{Julio A. Iglesias Martínez}
\affiliation{Institut FEMTO-ST, CNRS UMR 6174, University Bourgogne Franche-Comté, 15B Avenue des Montboucons, 25000 Besançon Cedex, France}

\author{Mahmoud Addouche}
\affiliation{Institut FEMTO-ST, CNRS UMR 6174, University Bourgogne Franche-Comté, 15B Avenue des Montboucons, 25000 Besançon Cedex, France}

\author{Ying Wu}
\email{ying.wu@kaust.edu.sa}
\affiliation{Computer, Electrical, and Mathematical Sciences and Engineering Division, King Abdullah University of Science and Technology (KAUST), Thuwal 23955-6900, Saudi Arabia}
\affiliation{Physical Science and Engineering (PSE) Division, King Abdullah University of Science and Technology (KAUST), Thuwal 23955-6900, Saudi Arabia}

\author{Abdelkrim Khelif}
\email{abdelkrim.khelif@femto-st.fr}
\affiliation{Institut FEMTO-ST, CNRS UMR 6174, University Bourgogne Franche-Comté, 15B Avenue des Montboucons, 25000 Besançon Cedex, France}
\affiliation{CINTRA, IRL 3288, CNRS/NTU/Thales, Research Techno Plaza, 50 Nanyang Drive, Singapore, 637553, Singapore}

\date{\today}

\begin{abstract}
The confinement of waves in open systems represents a fundamental phenomenon extensively explored across various branches of wave physics. Recently, significant attention has been directed towards bound states in the continuum (BIC), a class of modes that are trapped but do not decay in an otherwise unbounded continuum. Here, we theoretically investigate and experimentally demonstrate the existence of quasi-BIC (QBIC) for ultrasonic waves by leveraging an elastic Fabry-P\'erot metasurface resonator. We unveil several intriguing properties of the ultrasound QBIC that are robust to parameter scanning, and we present experimental evidence of a remarkable $Q$-factor of 350 at around 1 MHz frequency, far exceeding the state-of-the-art using a fully acoustic underwater system. Our findings contribute novel insights into the understanding of BIC for acoustic waves, offering a new paradigm for the design of efficient, ultra-high $Q$-factor ultrasound devices.
\end{abstract}

\maketitle

{ }

\clearpage


The pursuit of high quality-factor ($Q$-factor, or simply $Q$) resonators and systems has been a sustained endeavor spanning several years, driven by the potential of unlocking numerous compelling applications in acoustics \cite{kadic2013metamaterials,ma2016acoustic,cummer2016controlling}. Despite remarkable progress, various components, including acoustic sensors \cite{xu2010contour,farhat2021enhanced} for position and pressure, acoustic sources like sound 'lasers' \cite{farhat2021enhanced}, and acoustic transducers including microphones and loudspeakers, alongside numerous other acoustic devices \cite{assouar2018acoustic} still face challenges in achieving efficiently high-$Q$ resonant characteristics. This challenge primarily stems from the scarcity of acoustic resonators exhibiting $Q$-factors exceeding 100. To date, the $Q$-factors of resonances in underwater experimental setups have been confined to the range of a few tens \cite{biwa2005measurement,xu2010contour}, in a stark contrast to the capabilities observed in optics, where $Q$-factors can reach few millions \cite{ji2021methods} or even billions \cite{wu2020greater}. The complexity amplifies in the higher frequency regime, for example, ultrasound in the megahertz (MHz), and in aqueous environment where losses are further increased. 

Within the ultrasound spectrum, the $Q$-factor of a resonator encounters various constraints, with viscous damping in water playing a significant role. The presence of viscosity causes energy dissipation within the resonator \cite{korson1969viscosity,homentcovschi2008influence,morse1986theoretical}, leading to a reduction in the $Q$-factor. This phenomenon becomes more pronounced at higher frequencies and in confined spaces such as narrow channels \cite{ward2015boundary}. Additionally, acoustic radiation introduces another constraint, especially in the MHz range, as the resonators emit acoustic energy into the surrounding medium, particularly in open (non-Hermitian) environments \cite{huang2023acoustic}. This emission contributes to energy loss, further reducing the $Q$-factor. Material absorption, nonlinear effects, and temperature variations also negatively influence the confinement of the resonator. The combined impact of these factors poses significant challenges in achieving $Q$-factor beyond 100 at MHz frequencies in water \cite{pop2022lithium}. Overcoming these challenges is thus crucial for unlocking the full potential of high-$Q$ resonators in various applications, from advanced acoustic sensors to cutting-edge acoustic metamaterials. 

Recently, a new class of acoustic open resonators emerged, presenting the potential for achieving a high-$Q$ factors \cite{huang2021sound,huang2022general} through the utilization of bound states in the continuum(BIC). However, the focus of these studies has been on airborne acoustics at lower frequencies in the range of a few kilohertz (kHz). When transitioning to ultrasounds in the MHz range in water, the currently attainable $Q$-factors remain confined to the tens, primarily due to various influencing factors mentioned earlier \cite{hornig2022ultrasound}. 

 BIC exhibit a captivating phenomenon where wave modes are trapped in a specific region of space, even within an open system that permits energy to flow in and out \cite{hsu2016bound,joseph2021bound,koshelev2020engineering}. The emergence of BIC paved the way for the creation of resonant structures with high-$Q$ resonances \cite{hsu2016bound,joseph2021bound,koshelev2020engineering}. Originally discovered in quantum mechanics by von Neumann and Wigner, BIC results from the symmetry in the spatial distribution of the potential used to describe the wave equation, thereby creating a potential with localized eigen-fields at zero energy \cite{von1993merkwurdige}. In quantum mechanics \cite{landau2013quantum}, a BIC refers to a state of a particle that is confined to a potential well, demonstrating a tendency to remain localized in one region. It arises when the corresponding potential possesses a high degree of symmetry, interfering with outgoing waves to trap the mode within the energy continuum \cite{stillinger1975bound}. Consequently, this enables the mode to persist over an extended duration, despite its surrounding being an open system \cite{le2011quantum}.
 
 The application of BIC extends beyond quantum mechanics and has been since then observed in various physical systems, including optics \cite{hsu2013observation,marinica2008bound,bulgakov2008bound,plotnik2011experimental,monticone2014embedded,pankin2020one}, chiral metamaterials \cite{overvig2021chiral,zhou2023observation,chen2023observation}, elasticity \cite{lee2023elastic,marti2023bound}, and acoustics \cite{lyapina2015bound,huang2022general,amrani2021experimental,huang2021sound,huang2022topological,deriy2022bound,sadreev2022degenerate,chen2023phonon,liu2023observation,huang2023acoustic}. In particular, in acoustics, BIC can be created in resonant structures such as air cavities, narrow slits, and other systems that can partially confine sound waves or give rise to Fano resonances \cite{amin2015acoustically,amin2018perfect,amin2021polarization,bogdanov2019bound}. In general, an acoustic BIC may be attributed to two mechanisms: One involving the symmetry in the geometry of the resonant structure, known as symmetry-protected BIC resulted from the destructive interference between leaking waves \cite{lyapina2015bound,huang2022topological}. This mechanism is used to create ultra-high $Q$-factor resonances, leading to the development of devices such as filters, sensors, and resonant transducers \cite{huang2023acoustic}. The other mechanism involves mode coupling between resonant modes in different regions of the structure, which results in coupling-induced resonances. This can lead to either Fabry-P\'erot BIC (FP-BIC throughout this study) if the resonance frequencies of the cavities are equal \cite{hsu2016bound,amrani2021experimental} or Friedrich-Wintgen (FW-BIC) if the frequencies are different \cite{hsu2016bound,liu2023observation}.

Building upon this progress, our study aims to leverage the unique properties of BIC to create the highest $Q$-factor underwater ultrasound resonator. We present both theoretical analysis and experimental observations demonstrating that a double elastic metasurface, composed of an array of slits in a silicon plate, and immersed in water, can support FP-QBIC with an unprecedentedly high $Q$-factor. Experiments confirm the existence of a $Q$-factor on the order of 350, representing a substantial advancement beyond the current state-of-the-art in ultrasound resonators in a water environment.

\section*{Results}

\begin{figure}[t!]
    \centering
    \includegraphics[width=\columnwidth]{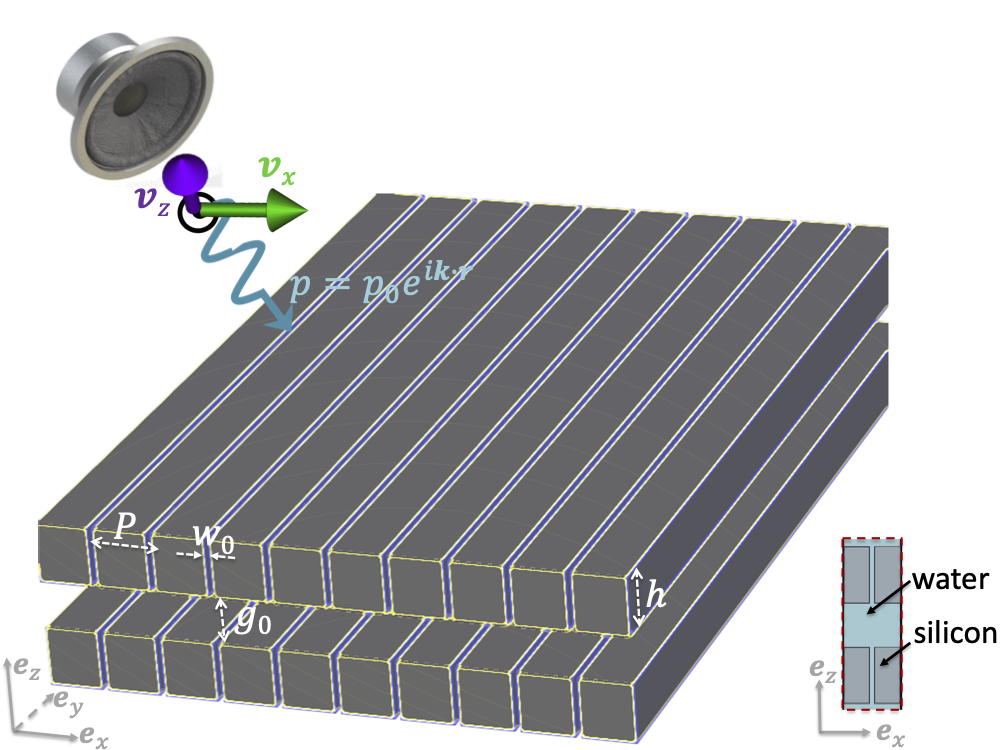}
    \caption{{\bf Scheme of the QBIC ultrasound metasurface.} Ultrasound waves in the frequency range of 0.5 to 2 MHz are generated (schematically) via the speaker at an oblique incidence on a periodic metasurface embedded in aqueous environment (water), which unit-cell, is depicted in the inset, with the corresponding geometry and material parameters.}
    \label{fig:fig1}
\end{figure}

\noindent {\bf Single unit-cell elastic metasurfaces}\\ The metasurface under consideration is schematized in Fig.~\ref{fig:fig1}. It comprises two parallel silicon slabs, each has 1 mm thickness in the $z$-direction, perforated with a periodic arrangement of thin slits aligned in the $x$-direction and extending infinitely in the $y$-direction (see inset of Fig.~\ref{fig:fig1} for the cross view). The silicon slabs are immersed in water. As a consequence, the slits are filled with water. Such configuration makes the problem numerically two-dimensional (2D). Each slab exhibits a periodicity of 1 mm. The slit, positioned at the center of the unit-cell (UC) has a width of 0.1 mm. In fact, the position of the slit in the $x$-direction is later demonstrated to have negligible influence for our purposes (see Supplementary Note 3). The medium surrounding the UC (colored in cyan) shown in the inset of Fig.~\ref{fig:fig1} is water, including the space inside the slit. 

We use the finite-element software COMSOL Multiphysics \cite{comsol} to characterize the scattering response (transmission/reflection spectrum) of this metasurface (a single silicon slab) for ultrasound frequencies ranging from 500 to 1500 kHz. The propagation of sound waves in water is described via pressure acoustic module and that in silicon via solid mechanics module, as silicon supports both longitudinal and shear waves propagation (see Supplementary Note 1). 

\begin{figure*}
    \centering
    \includegraphics[width=1.6\columnwidth]{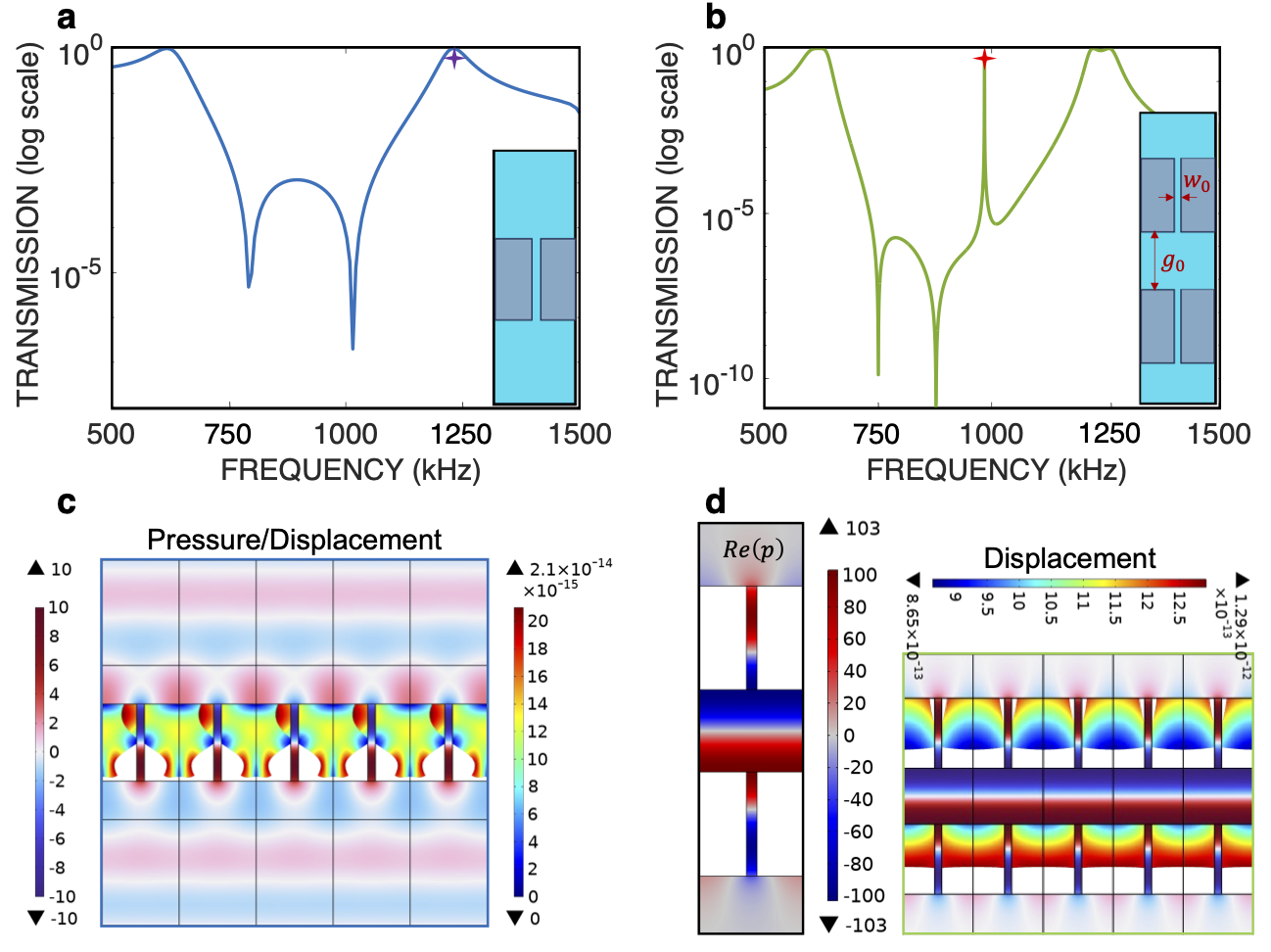}
    \caption{{\bf Transmission characteristics of the FP-based QBIC cavity.} {\bf a} Transmission in logarithmic scale of the single acoustic metasurface (shown in the inset). {\bf b} Transmission for the double metasurface (QBIC resonator, shown in the inset) in logarithmic scale. {\bf c} Real part of the pressure field (left color bar) and displacement amplitude (right color bar) at the frequency 1231.1 kHz, corresponding to the second peak,  highlighted by a purple star in Fig.~\ref{fig:fig2}{\bf a}. {\bf d} Real part of the total pressure field at the QBIC point (left panel), highlighted by a red star in Fig.~\ref{fig:fig2}{\bf b} and amplitude of the displacement field at the same frequency (right panel).}
    \label{fig:fig2}
\end{figure*}

Figure~\ref{fig:fig2}(a) presents the transmittance ($|t|^2$) spectrum of this metasurface in logarithmic scale under normal incidence, exhibiting some regular FP resonances around 600 kHz and 1200 kHz and two dips resembling a $W$-shape around 830 kHz and 1000 kHz. It should be emphasized that these resonances are broadband. For instance, the $Q$-factor of a resonance is defined as $Q=\omega_0/\Delta\omega$, where $\omega_0$ is the resonance frequency and $\Delta\omega$ is the full-width-at-half-maximum (FWHM) or alternatively and equivalently as $Q=\omega_0/2\Gamma$, with $\Gamma$ being the imaginary part of the complex frequency $\omega=\omega_0-i\Gamma$ that represents the radiative decay rate of the leaky mode. The resonances shown in Fig.~\ref{fig:fig2}(a) have a low $Q$-factor of around 10. In particular, the resonance highlighted by a purple star has a near-field of pressure, depicted in Fig.~\ref{fig:fig2}(c) that is not perfectly localized within the metasurface, indicating high leakage or $\Gamma\approx\omega_0$. The amplitude of pressure (its real part) and the displacement field $|\Re{({\bf w})}|$ in the solid silicon layers are comparatively low, which is a characteristics of an overdamped acoustic resonance. Hence, this single metasurface cannot be used alone for the applications that we envision, e.g., sensing or acoustic lasing systems that require a $Q$-factor of the order of a few hundreds, which is still lacking for underwater ultrasound waves.\\

\noindent {\bf Double unit-cell elastic metasurfaces and ultrasound QBIC}\\ 
To address the challenges of low $Q$-factors at ultrasound frequencies, where losses are prominent, we leverage the concept of BIC. This phenomenon is known to yield diverging $Q$-factors for certain modes embedded in the continuum of radiating modes. To exploit this effect, we utilize the same metasurface depicted in the inset of Fig.~\ref{fig:fig2}(a), by cascading two such metasurfaces, as illustrated in Fig.~\ref{fig:fig2}(b). This configuration is designed to potentially trap acoustic radiation in the gap between the two metasurfaces, leading to the emergence of BIC under specific conditions. For instance, the BIC takes place when the gap thickness $g_0$ is chosen in such a way that the accumulated phase in a single round-trip equals  $m\times2\pi$, with $m\in\mathbb{N}$. This principle reflects the analogy with Fabry-P\'erot cavities, typically composed of two mirrors. It is worthy to emphasize the importance of using a solid as the building block for the metasurface to allow the generation of BIC. If a hard-wall is used instead, the QBIC disappears as shown in the Supplementary Note 2 (considering oblique incidence in these cases). This underscores the unique contribution of the solid-based metasurface to create and sustain the QBIC in the ultrasound regime. 

To gain insight of this FP-BIC, we employ the temporal coupled-mode theory (TCMT) \cite{fan2003temporal,verslegers2010temporal,overvig2023spatio}. We denote the amplitude of the resonances as a state vector $\Psi=(\psi_1,\psi_2)^T$ that obeys the governing equation $\hat{H}\Psi=i\partial\Psi/\partial t$, with the Hamiltonian operator written as:
\begin{equation}
\hat{H}=
\begin{pmatrix}
\omega_0 & \delta\\
\delta & \omega_0
\end{pmatrix}-i\Gamma
\begin{pmatrix}
1 & e^{i\varphi}\\
e^{i\varphi} & 1
\end{pmatrix}\, .
\label{eq:eq-Hamil}    
\end{equation}
Here, $\omega_0$ is the resonance frequency of the single metasurface, highlighted by the purple star in Fig.~\ref{fig:fig2}(a), which is the same for both metasurfaces. It is crucial to note that if the two metasurfaces possess distinct resonant frequencies, $\omega_1$ and $\omega_2$ for instance, the resulting BIC will be fundamentally different, i.e., a FW-BIC \cite{hsu2016bound}. $\delta$ represents the near-field coupling between the two metasurfaces, a parameter strongly dependent on the gap $g_0$ as can be seen in the last sub-section of Supplementary Note 3, and with the the frequency $\omega_1$ and the QBIC frequency, we have $\delta=1.55\times10^6$ rad/s ($2\pi\times 246.7$ kHz). $\Gamma=1.98\times10^5$ rad/s ($2\pi\times 31.5$ kHz) is the radiative decay rate of the single metasurface and $\varphi=kg_0$ is the phase-shift between the resonators, where $k$ denotes the acoustic (pressure) wavenumber. This system possesses two eigen-solutions $\Psi_\pm$, represented by the eigen-values of Eq.~(\ref{eq:eq-Hamil}):
\begin{equation}
\omega_\pm=\omega_0\pm\delta-i\Gamma\left(1\pm e^{i\varphi}\right) \, .
\label{eq:eq-eigen}    
\end{equation}
From Eq.~(\ref{eq:eq-eigen}) it is clear that when $\varphi=m\times\pi$, meaning the phase accumulation along one round-trip is $m\times2\pi$, the eigen-frequencies of the Hamiltonian simplify to $\omega_\pm=\omega_0\pm\delta-i\Gamma(1\pm(-1)^m)$. Hence, one of the modes ($\Psi_-$) exhibits a zero imaginary part, while the other ($\Psi_+$) possesses an imaginary part of $2\Gamma$, i.e., a radiative decay rate double that of the single metasurface. The occurrence of a purely-real eigen-frequency characterizes an authentic BIC.

\begin{figure*}
    \centering
    \includegraphics[width=1.6\columnwidth]{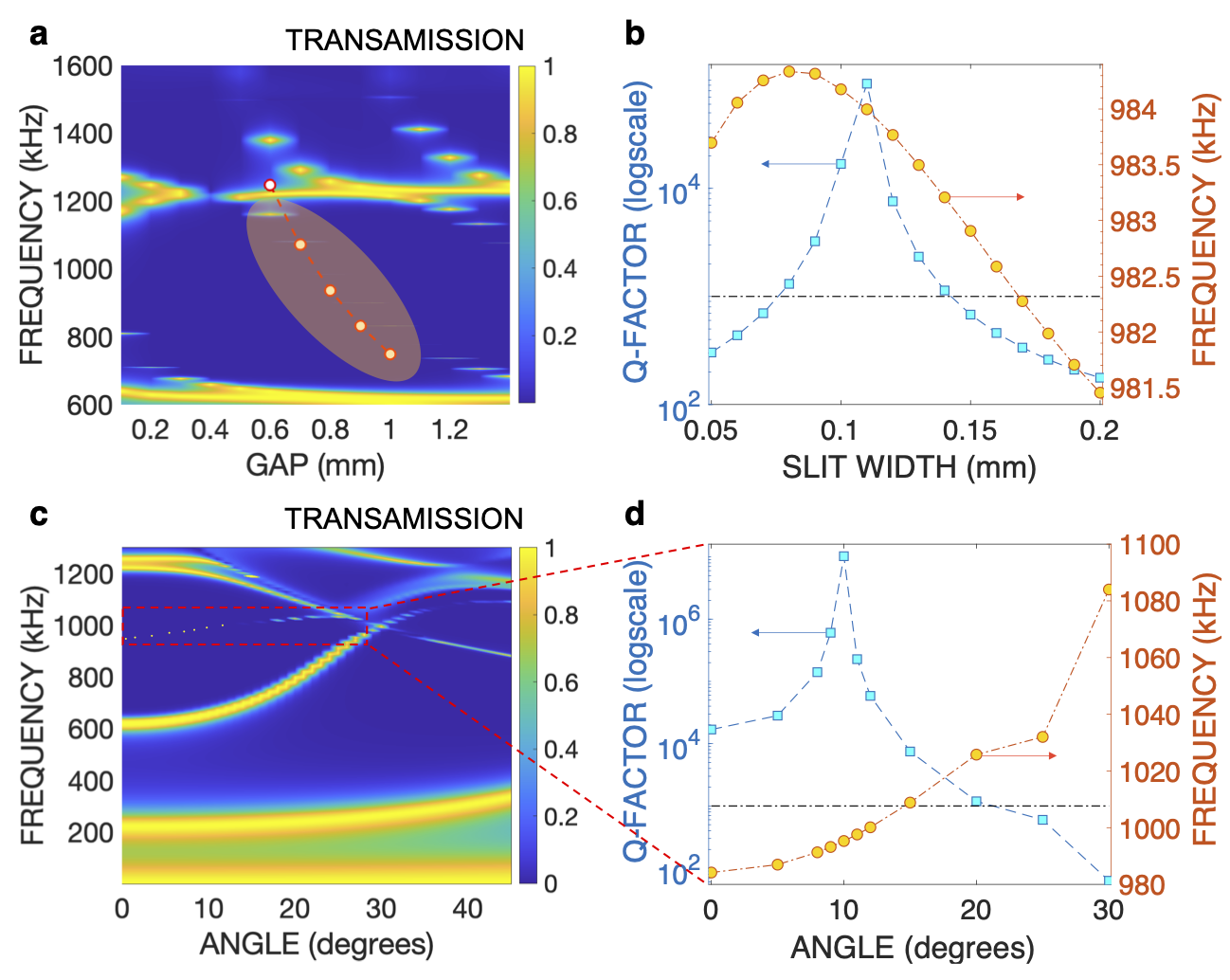}
    \caption{{\bf Effect of the gap and slit widths and dispersive properties.} {\bf a} Contour plot of the transmission of the metasurface of same parameters as in Fig.~\ref{fig:fig2}{\bf b} VS gap $g_0$ and frequency. The white dotted curve depicts the TCMT predictions and the highlighted region shows the QBIC regime. {\bf b} $Q$-factor for varying slit width $w_0$ for a gap of 0.8 mm. The right $y$-axis gives the frequency of the corresponding resonant QBIC. {\bf c} Dispersion curve of the BIC, i.e., versus incidence angle from 0 to $45^{\circ}$. The dashed red square denotes the location of the BIC, that is plotted in {\bf d} with its corresponding $Q$-factor and resonance frequency (left and right $y$-axis, respectively).}
    \label{fig:fig3}
\end{figure*}

Figure~\ref{fig:fig2}(b) illustrates the numerically calculated transmission of a double metasurface sharing identical parameters with those in Fig.~\ref{fig:fig2}(a), albeit featuring a gap of $g_0=0.8$ mm (Referring to Supplementary Note 3 for confirmation that the position of the slit in the $x$-direction does not affect our results). In addition to the previously observed maxima and minima resembling a $W$-shape with the single metasurface, a distinct ultra-narrow peak emerges at $f_0=984.5$ kHz in the double metasurface scenario. The distinguishing feature signaling the nature of this resonance as a QBIC is primarily its remarkably high $Q$-factor, estimated here at around 16,000, a number of 3 orders of magnitude higher than the $Q$-factor associated with the single metasurface design. Furthermore, the observation of Fig.~\ref{fig:fig2}(d) with both the real part of the pressure field (left panel) and the displacement field in elastic solid (right panel) shows the strong confinement of acoustic waves around the metasurface. The acoustic energy attains exceptionally high values in the gap and the slits, reaching around $10^4$, and similarly the elastic displacement is two orders of magnitude higher compared to the single metasurface case. In addition, the QBIC frequency predicted by the simplified TCMT gives a value of 935 kHz, closely aligning (within a 4$\%$ error margin) with the observed frequency in Fig.~\ref{fig:fig2}(b), denoted by the red star. The discrepancy between these two values may be attributed to the metasurface sustaining both shear and longitudinal waves \cite{graff2012wave}, so that a more adapted (though complicated) modeling is desired for a precise alignment with the simulations.


\noindent {\bf Tunability and dispersion of the QBIC}\\ 
The frequency of the proposed FP-QBIC is tunable. Beginning with the same metasurfaces as previously detailed in Fig.~\ref{fig:fig2}(b), we change the gap $g_0$ between the two metasurfaces. The transmission data is plotted in Fig.~\ref{fig:fig3}(a), with the $x$-axis representing $g_0$ in mm and the $y$-axis representing frequency in kHz. The highlighted zone indicates the dependence of resonance frequency$f_0$ on $g_0$, agreeing well with analytical predictions from TCMT depicted by the white dotted curve in Fig.~\ref{fig:fig3}(a) (additional details are provided in the first sub-section of Supplementary Note 3). It is noteworthy that the QBIC disappears for smaller gaps and the $Q$-factor reaches its optimum value for $g_0\approx0.8$ mm. Hence, we select $g_0=0.8$ mm for experimental validation detailed in the next sub-section. We also analyze the impact on the $Q$-factor of altering the slit width $w_0$ in Fig.~\ref{fig:fig3}(b). For small values of $w_0$, such as approximately 50 $\mu$m, the $Q$-factor deteriorates. An optimal value is obtained for $w_0=0.11$ mm, at which the $Q$-factor reaches approximately $10^5$. As $w_0$ increases, the $Q$-factor drops again, reinforcing that the most favorable BIC configuration corresponds to $w_0\approx0.1 P$, i.e., 10$\%$ of the period of the metasurface. The corresponding change of the resonance frequency is also detailed in Fig.~\ref{fig:fig3}(b) (additional details are provided in Supplementary Note 3).

\begin{figure*}
    \centering
    \includegraphics[width=1.6\columnwidth]{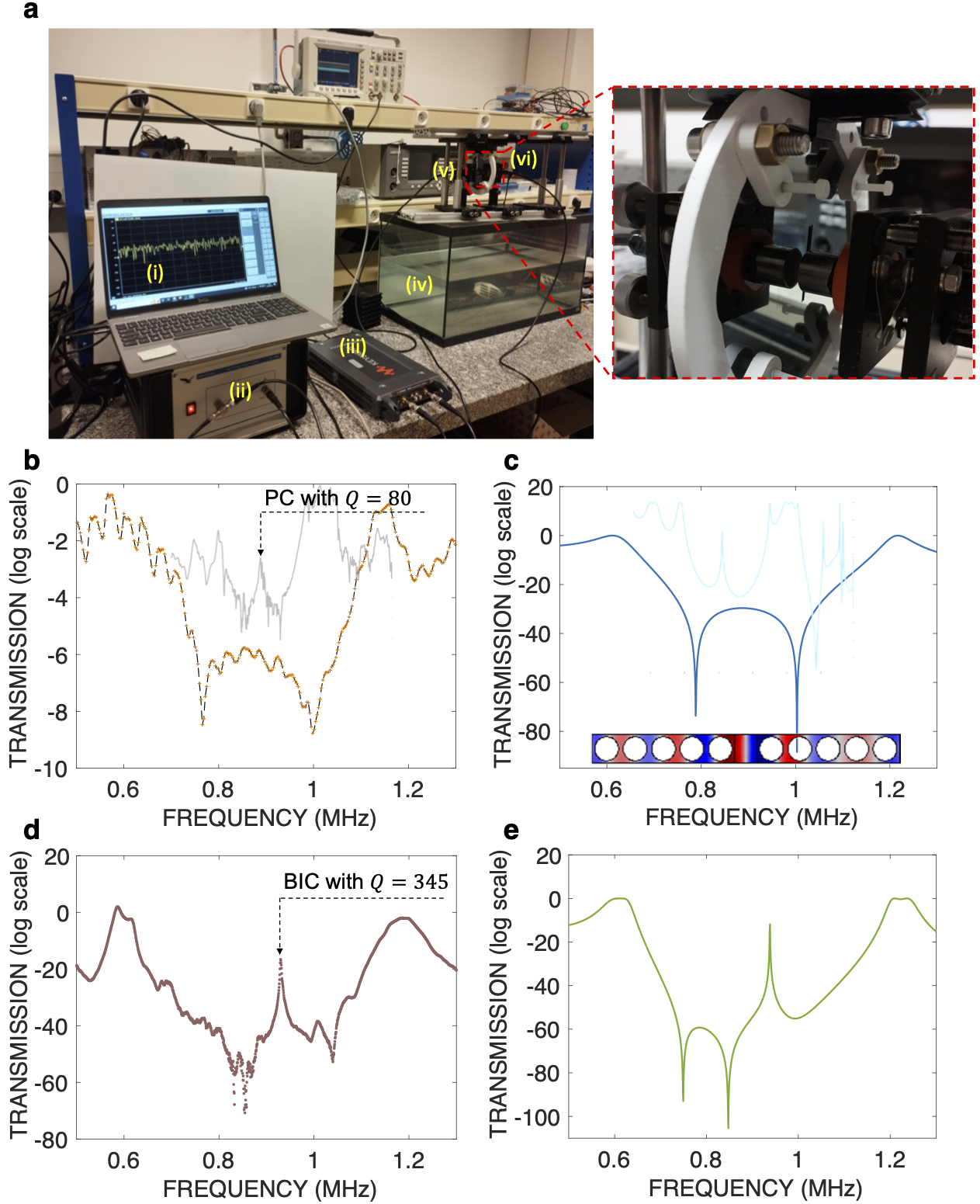}
    \caption{{\bf Experimental results.} {\bf a} Experimental setup used to characterize the transmission/reflection spectrum of the metasurfaces in the ultrasound regime underwater: (i) Ultrasound signal generation and data acquisition system; (ii) Amplifier; (iii) Vector network analyzer (VNA); (iv) Water reservoir; (v)-(vi) Basic mount with a rotating element (zoomed in in the inset). {\bf b} Experimental and {\bf c} numerical modeling of the transmission of the single silicon metasurface (log scale). The insets (transparent plots) give the transmission from a phononic cavity both numerically and experimentally using steel rods of diameter 2 mm, period of 2.22 mm and cavity size of 1.78 mm (See Supplementary Note 4 for details). {\bf d} Experimental and {\bf e} numerical modeling of the transmission of the double silicon metasurface (log scale). The measured $Q$-factor of the BIC is 345.}
    \label{fig:fig4}
\end{figure*}

It is also important to investigate the dispersive properties, specifically the angular dependence, of the QBIC mechanism. To accomplish this, we vary the incident angle of the incoming acoustic plane wave in the range 0-45$^{\circ}$ and plot the resulting transmission in a 2D format in Fig.~\ref{fig:fig3}(c). The regular broadband resonances, attributed to the FP interference, coexist with the QBIC (evidenced by a vanishing line-width) as highlighted by the red-dashed square. Figure~\ref{fig:fig3}(d) shows the dependence of the $Q$-factors on incident angles, up to 30$^{\circ}$. Notably, the $Q$-factor demonstrates an increase from its value at normal incidence ($\approx10^4$) with ascending angles, peaking at $\approx10^7$ for a 10$^{\circ}$ incidence, before decreasing. For angles exceeding 20$^{\circ}$, the $Q$-factor drops significantly below the threshold depicted by the dotted-dashed-black line corresponding to $Q=10^3$, indicating that the resonance can no longer be considered as a BIC. The notable increase in $Q$-factor with angle and its optimum value around 10$^{\circ}$ (i.e., off-Gamma point) is of accidental nature, and is thus reminiscent of accidental BIC. The resonance frequency of the BIC versus angle is illustrated by the right $y$-axis of Fig.~\ref{fig:fig3}(d), and it consistently increases with angle within the considered domain.\\

\noindent {\bf Experimental demonstration of FP-BIC}\\ We fabricate two metasurfaces made of silicon material with identical geometric properties as analyzed in previous sections. Our experimental setup involves employing a dicing machine to create periodic perforations in two identical silicon wafers, as illustrated in Fig.~\ref{fig:fig4}(a). Each silicon wafer possesses a thickness of $1$ mm. The slits are positioned at intervals defined by a pitch $P$ of 1 mm, each slit being 0.1 times the pitch in width, inducing $w_0=0.1$ mm. This configuration covers an area of 33 by 40 square millimeters on the wafer's surface, equivalent to 33 periods and slits measuring 40 mm in length. To maintain  a consistent cavity gap of $g_0=0.8$ mm, a plastic ring is inserted between the two wafers.  

Figures~\ref{fig:fig4}(b) and~\ref{fig:fig4}(d) depict the transmission (log scale) of the single and double metasurface, respectively, for frequencies between 0.5 and 1.3 MHz. Excellent agreement is found between experimental [Figs.~\ref{fig:fig4}(b) and~\ref{fig:fig4}(d)] and numerical results [Figs.~\ref{fig:fig4}(c) and~\ref{fig:fig4}(e)]. Particularly, the resonant frequency found experimentally corresponds to a gap of 0.85 mm, falling within the experimental uncertainty related to fabrication and measurement errors (up to 4 $\%$). The experimentally obtained $Q$-factor, as illustrated in Fig.~\ref{fig:fig4}(d) reaches 345 (refer to Supplementary Note 4 for further experimental details). Even though this experimentally observed BIC is orders of magnitude lower than its numerical counterpart (approximately $10^4$), it still offers a significant advancement for underwater ultrasound resonators. To the best of our knowledge, resonators exhibiting similar $Q$-factors in underwater ultrasound applications have not been reported so far. In fact, the highest observed $Q$-factors in this regime were achieved using phononic cavities, such as defects in 1D arrays of solid rods, for example, steel. To highlight the superior performance of our BIC design, we construct a phononic cavity using steel rods with diameter of 2 mm, period of 2.22 mm, comprising 6 rows and a central gap of 1.78 mm, as shown in Figs.~\ref{fig:fig4}(b)-(c). We perform both numerical calculation and experimental measurement of transmission of such resonator. The maximum $Q$-factor that we can measure from such structure is around 80. Hence, our FP-BIC resonator demonstrates a substantially higher $Q$-factor, providing valuable potential for future investigations in ultrasound sensors, acoustic imaging systems, and acoustic metasurfaces in general. 

\section*{Discussion}

\begin{figure}[t!]
    \centering
    \includegraphics[width=0.8\columnwidth]{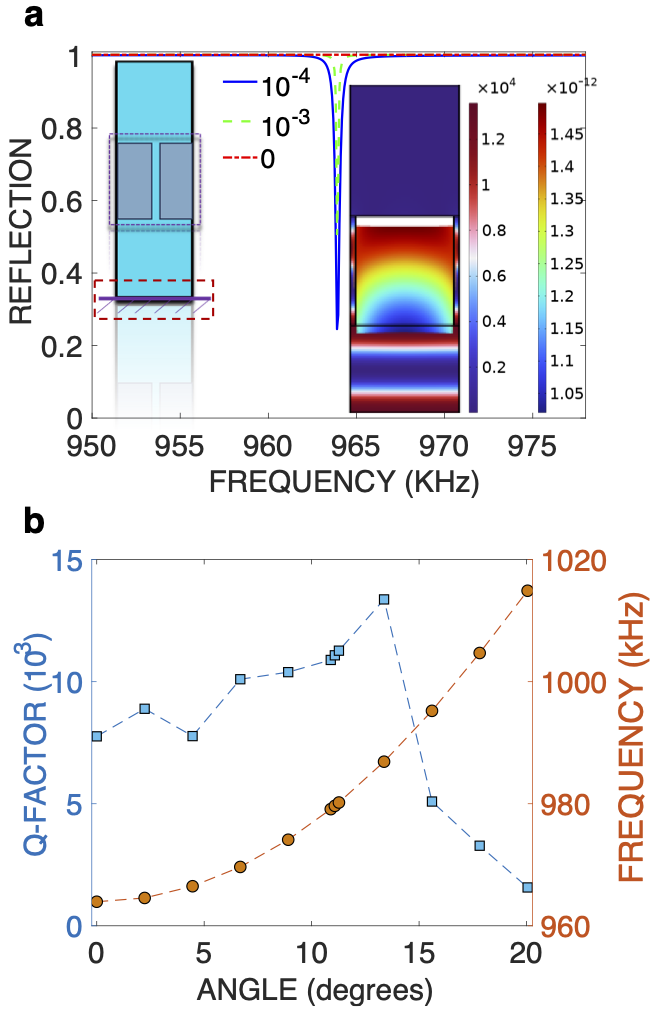}
    \caption{{\bf Single-UC QBIC.} {\bf a} Reflection spectrum for the single metasurface QBIC, schematized in the left inset, where a mirror is present at distance $g_0$. The right inset depicts the real part of the pressure and displacement for that QBIC around 965 kHz. {\bf b} $Q$-factor (in units of $10^3$) of the obtained QBIC VS the angle of incidence (left axis: blue curve) and resonance frequency in kHz VS the angle of incidence (right axis: red curve).}
    \label{fig:fig5}
\end{figure}

We have demonstrated in Eq.~(\ref{eq:eq-Hamil}) that the formation of ultrasound QBIC necessitates two \textit{elastic} metasurfaces for both the FP and FW QBIC. However, a workaround to this limitation emerges by introducing a mirror (perfectly-reflecting surface) beneath the single metasurface.  We show that the FP-QBIC can be observed under such conditions.  When only a single metasurface is considered, the FP-BIC disappears as the interference phenomenon is vital to QBIC formation [See Eq.~(\ref{eq:eq-eigen})]. In pressure acoustics, a perfect mirror is realized by enforcing sound hard-wall boundary condition (${\bf n}\cdot\frac{1}{\rho}\nabla p$) \cite{farhat2020scattering}. The mirror, acting as the second resonator, ultimately leads to the QBIC, observed in Fig.~\ref{fig:fig5}(a) at a resonant frequency of 964 kHz, which is even closer to the TCMT predicted value. Importantly, to observe this QBIC, we need to add some loss in the narrow water channels; otherwise all the energy would be reflected [$|R|=1$, in a lossless scenario, as depicted by the red dashed line in Fig.~\ref{fig:fig5}(a)]. The plots in Fig.~\ref{fig:fig5}(a) showcase increasing losses (imaginary part in the speed of sound in water). The near-field plots of both pressure and displacement are shown in the inset of Fig.~\ref{fig:fig5}(a), right panel, revealing a comparable enhancement to the double-metasurface scenario. Of course, the quality of the mirror and its practical realization are pivotal to the QBIC properties and $Q$-factor; these aspects will be further investigated in future studies.

In Figure~\ref{fig:fig5}(b), the variation of both the $Q$-factor and the resonance frequency of the QBIC in single-metasurface-with-mirror when the angle of incidence of the wave varies between 0 and 20$^{\circ}$. It is noticed that the $Q$-factor increases with oblique incidence angle and reaches a maximum around 14$^{\circ}$, and then starts decreasing. This behavior bears some similarity to the case of the double-metasurface QBIC shown in Fig.~\ref{fig:fig3}(d), albeit with a much smaller value of $Q$, potentially attributed to symmetry breaking in case of the hard-wall mirror.

In summary, our work introduces a theoretical framework for ultra-high $Q$-factor ultrasound resonators designed for underwater acoustic applications. Experimental validation confirms the achievement of the targeted high-$Q$, utilizing innovative design strategies leveraging the concept of QBIC with either two elastic metasurfaces or a single metasurface combined with a mirror. This approach exploits the destructive interference phenomenon to achieve unprecedented $Q$-factors for the underwater ultrasound waves. The exceptional $Q$-factors associated with FP resonances are intricately linked to the waveguide perturbation, offering a versatile means of attaining ultrahigh-$Q$ resonances without necessitating topological concepts, as exemplified in recent literature. What distinguishes our proposed approach is its potential for achieving ultrahigh $Q$-factors while avoiding complicated processes that often introduce sample roughness and disorder. Our experimental demonstration, conducted at a wavelength of approximately 1 MHz using a thin silicon-slitted metasurface slab, yields remarkable $Q$-factors of up to 350, rivaling those achieved through topological charge engineering. Furthermore, we achieve tunability of the resonant wavelength by varying the slit width, gap size, and/or angle of incidence. 

Additionally, by generalizing our structure to 3D geometry, i.e., selecting distinct lattice constants along the $x$- and $y$-axes, we can enhance the precision of realistic QBIC resonators. These findings represent a promising avenue for the development of ultrasonic and acoustic high-$Q$ devices, offering the potential to enhance performance in applications such as biosensors \cite{zhang2021quasi}, reflectors \cite{luo2022wavy}, and filters through the utilization of ultra-sharp spectral features.

\section*{Methods}

\noindent {\bf Numerical simulations}\\ The numerical simulations are performed via the commercial software COMSOL Multiphysics \cite{comsol}. Sound waves propagating in water are modeled via the 'Pressure Acoustics, Frequency Domain (acpr)'. Elastic waves propagating in solid silicon are modeled via the 'Solid Mechanics (solid)'. At the interface, multiphysics 'Acoustic-Structure Boundary' conditions are used. In addition, to excite the structure, for any angle of incidence, we use background pressure field, where we define an impinging plane acoustic wave of wave-vector ${\bf k}=(k_x,k_y)$, and a given pressure amplitude and phase, taken as 1 and 0, respectively. The top and bottom domains are taken as perfectly matched layers (PMLs) to avoid domain reflections. The lateral boundary conditions are Floquet pseudo-periodic with a ${\bf k}$-vector $(k_x,k_y)$, both in the acpr and solid domains. The transmittance and reflectance are calculated by integration over a boundary on top and bottom of the metasurfaces, respectively.\\

\noindent {\bf Metasurface fabrication and measurement}\\ The silicon utilized in this experiment is of the <100> crystallographic orientation, characterized by a mass density of 2329 kg/m3. Along the [100] direction, it exhibits longitudinal and transverse wave velocities of 8433 m/s and 5843 m/s, respectively.
The experimental procedure relies on ultrasound transmission technique, wherein the sample is immersed in water and positioned between the generating and receiving transducers. These transducers, with a diameter of 15 mm, are broadband, centered at a frequency of 1 MHz frequency. A vector network analyzer (VNA) is employed to generate a wide-ranging signal centered precisely at 1 MHz. This signal is then amplified to ensure a consistent power level for activating the transducer. The resulting pulse is received by a second transducer, linked to the second port of the VNA, which can perform a narrow-band scan across a frequency range. At each frequency point, both the amplitude and phase of the signal are recorded. To establish a baseline, the transmission spectra are normalized by calculating the ratio between spectra obtained with and without the sample in place.
Furthermore, an additional evaluation was conducted on a homogeneous membrane, classical FP resonator, as well as a phononic crystal to assess the impact of the quasi-BIC confinement.

\section*{Data availability}

All data are available in the main text or the Supplementary materials.

\section*{Code availability}

The code used for the analyses will be made available upon reasonable e-mail request to the corresponding authors.


\providecommand{\noopsort}[1]{}\providecommand{\singleletter}[1]{#1}%

\begin{acknowledgments}
M.F. and Y.W. acknowledge support by King Abdullah University of Science and Technology (KAUST) Office of Sponsored Research (OSR) under Grant No. ORFS-CRG11-2022-5055, as well as KAUST Baseline Research Fund BAS/1/1626-01-01. J.I.M., M.A., and A.K. acknowledge the financial support from the Region of Franche-Comte under the reference 2012C-08901, as well as the support by the french RENATECH network and its FEMTO-ST technological facility.
\end{acknowledgments}

\section*{Author contributions}

A.K. and Y.W. conceived the idea; M.F. and Y.A. performed the numerical simulations and generated the data, with help from M.A.; Y.A., J.I.M., M.A., and A.K. conducted the experimental fabrication and measurement; All authors did the physical analysis and contributed to the writing of the manuscript, the first draft of which was written by M.F.; Y.W. and A.K. supervised the project. 

\section*{Competing interests}

The authors declare no competing interests.

\section*{Additional information}

\noindent {\bf Supplementary information} The online version contains supplementary material available at https://doi.org/xxxx

\noindent {\bf Correspondence} and requests for materials should be addressed to Ying Wu and Abdelkrim Khelif.

\noindent {\bf Reprints and permission information} is available at http://www.nature.com/reprints

\noindent {\bf Publisher’s note} Springer Nature remains neutral with regard to jurisdictional claims in published maps and institutional affiliations

\noindent {\bf Open Access} This article is licensed under a Creative Commons Attribution 4.0 International License, which permits use, sharing, adaptation, distribution and reproduction in any medium or format, as long as you give appropriate credit to the original author(s) and the source, provide a link to the Creative Commons licence, and indicate if changes were made. The images or other third party material in this article are included in the article’s Creative Commons licence, unless indicated otherwise in a credit line to the material. If material is not included in the article’s Creative Commons licence and your intended use is not permitted by statutory regulation or exceeds the permitted use, you will need to obtain permission directly from the copyright holder. To view a copy of this licence, visit http://creativecommons.org/ licenses/by/4.0/.

\end{document}